\documentclass{appolb}
\usepackage{graphicx}
\usepackage{xcolor}
\usepackage{float}
\usepackage{amsmath}

\def\Pom{{\bf I\!P}}
\def\Reg{{\bf I\!R}}

\begin{document}
\title{Phenomenology developments in UPC: 
$\gamma \gamma \to \gamma \gamma$ scattering.%
\thanks{Presented at XXXII Cracow Epiphany Conference on the recent results from Heavy Ion Physics}%
}
\author{Paweł Jucha, Antoni Szczurek
\address{Institute of Nuclear Physics PAN, ul. Radzikowskiego 152, 31-342 Kraków, Poland}
}
\maketitle
\begin{abstract}
We discuss several possible extensions of present studies
of $\gamma \gamma \to \gamma \gamma$ scattering in ultraperipheral
heavy ion collisions. One of the possible extensions are studies
for lower diphoton invariant masses (smaller transverse momenta). 
There new mechanisms may show up.
This includes possible studies with future FOCAL and ALICE 3 detectors.
So far only coherent $\gamma \gamma \to \gamma \gamma$ processes,
when initial photons couple to nuclei, were sudied theoretically. 
Recently we proposed to study also inelastic processes, when initial 
photons couple to individual nucleons. 
Without special cuts the corresponding processes are
of the order of 20 - 30 \%. A study to which extent the inelastic 
processes survive with the present cuts used for the 
$\gamma \gamma \to \gamma \gamma$ studies is an interesting question. 
They can be of interest by itself and dedicated studies are suggested.
We calculate also cross section for associated neutron production. Deviation
from our predictions could signal presence of the inelastic contribution.
Finally we discuss production of single photons in UPC 
($A A \to A A \gamma$). Corresponding cross sections for selected mechanisms  are 
presented.
\end{abstract}
  
\section{Introduction}

Almost 10 years passed since first experimental observation of 
light-by-light scattering in lead-lead ultraperipheral collisions 
(UPC) was made by the ATLAS Collaboration \cite{ATLAS}. At the end of 
this decade, we present a summary of possible extensions, which 
could bring new perspective on the topic.
Despite precise measurements of ATLAS \cite{ATLAS:2020hii} and CMS experiments~\cite{CMS}, the 
discrepancy between theory and data are still not fully resolved issue.
Improvements in mathematical apparatus, calculation of next-to-leading
order corrections \cite{{AH:2023ewe},AH:2023kor}, 
or proposal of different mechanisms did not fill-up the gap. 
Thus, we propose the expansion of the research to new regions.
With development of detectors like ALICE FoCal and ALICE 3 the cut on 
photon energy will be lowered \cite{JKS2024}, which will increase 
the statistics and potentially allow the observation of new mechanisms 
involved in 
$\gamma \gamma \to \gamma \gamma$ scattering process.
We also discuss the impact of inelastic light-by-light scattering, 
which was not considered before. The correlation between
nuclear disintegration and $\gamma \gamma \to \gamma \gamma$ 
scattering may help in identification of
inelastic processes. Finally, we propose a measurement
of single photon, where photo-production with signal from 
one $\gamma$ need to be  observed.

\section{Theoretical calculation of nuclear
	cross sections}

The general formula for nuclear cross section is calculated using 
equivalent photon approximation (EPA) in the impact parameter space. 
In this well-established approach, the di-photon cross section can 
be written as:
\begin{eqnarray}
	\frac{d\sigma(PbPb \to PbPb \gamma \gamma)}{dy_{\gamma_1}dy_{\gamma_2}dp_{t,\gamma}} &=& 
	\int
	\frac{d\sigma_{\gamma\gamma\to\gamma\gamma}(W_{\gamma\gamma})}{dz}
	N(\omega_1,b_1) N(\omega_2,b_2) S^2_{abs}(b) \nonumber \\  
	&\times& d^2b d\bar{b}_x d\bar{b}_y \frac{W_{\gamma\gamma}}{2} \frac{dW_{\gamma\gamma}dY_{\gamma\gamma}}{dy_{\gamma_1}dy_{\gamma_2}dp_{t,\gamma}} dz \;,
	\label{eq:tot_xsec}
\end{eqnarray}
\noindent where $\bar{b}_x = \left( b_{1x} + b_{2x}\right)/2$ and $\bar{b}_y = \left( b_{1y} + b_{2y}\right)/2$. The relation between vectors $\vec{b}_1$, $\vec{b}_2$ and impact parameter is $b = |\vec{b}| = \sqrt{|\vec{b}_1|^2 + |\vec{b}_2|^2 - 2|\vec{b}_1||\vec{b}_2|\cos\phi}$. The nuclear photon fluxes 
$N(\omega_1,b_1)$ and $N(\omega_2,b_2)$ are calculated usually with realistic charge distribution. In our paper \cite{JKS2024}, we discussed the impact
of different survival factors $S^2_{abs}(b)$. The sharp edge of nucleus can be implemented as:
\begin{eqnarray}
	 S^2_{abs}(b) = \Theta(b-b_{max})  .
\end{eqnarray}
\noindent The second approach includes the nuclear surface smearing, 
described mathematically by the formula:  
\begin{equation}
	 S^2_{abs}(b) = \text{exp}\left( -\sigma_{NN} T_{AA}(b) \right)  \;,
	\label{eq:s2b}
\end{equation}
\noindent where $\sigma_{NN}$ is the nucleon-nucleon interaction cross
section.
$T_{AA}(b)$ above is related to the so-called nuclear thickness, 
$T_A(b)$,
\begin{equation}
	T_{AA}\left(|\vec{b}| \right) = \int d^2\rho 
	T_A \left( \vec{\rho}-\vec{b} \right) T_A \left( \vec{\rho} \right) ,
\end{equation}
\noindent and the nuclear thickness is obtained by integrating the nuclear density 
\begin{equation}
	T_A \left( \vec{\rho} \right) = \int \rho_A\left( \vec{r} \right) dz  ,
	\hspace{0.5cm} \vec{r} = \left( \vec{\rho},z \right) \;,
\end{equation}
\noindent where $\rho_A$ is the nuclear charge distribution described 
by the Fermi parametrization.

\begin{figure}
	a)\includegraphics[scale=0.4]{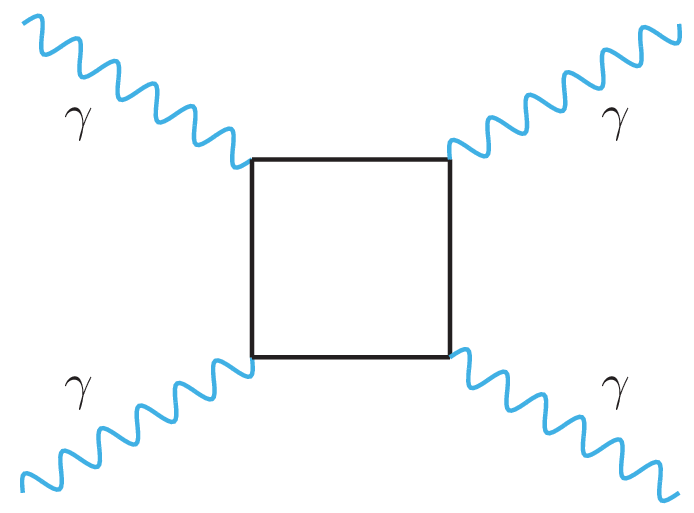} \hspace{0.75cm}
	b)\includegraphics[scale=0.4]{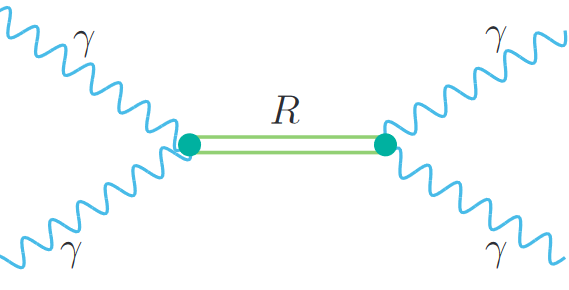} \\
	c)\includegraphics[scale=0.45]{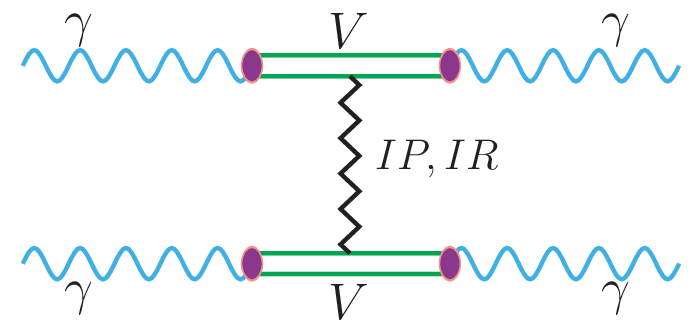} \hspace{0.25cm}
	d)\includegraphics[scale=0.45]{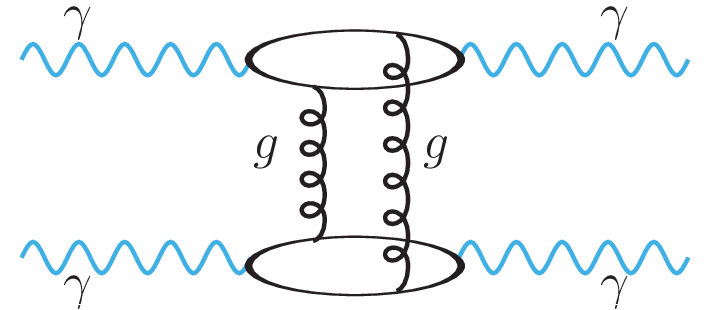}
	\caption{Possible mechanisms for 
        $\gamma \gamma \to \gamma \gamma$ scattering: 
        a) fermionic box, b) light-meson resonances
	c) VDM-Regge vector mesons fluctuation, d) 2-gluon exchange.}
	\label{fig:diagrams}
\end{figure}

In Fig.\ref{fig:diagrams} we show mechanisms included
by our group. We will not discuss them in more detail here,
except of double vector meson fluctuations which is considered shortly
in the next section.

\section{Double vector meson fluctuations}

The third component presented in Fig. \ref{fig:diagrams}c), was 
calculated for the first time by Klusek-Gawenda, Lebiedowicz, 
Szczurek in \cite{KLS2016}, assuming vector dominance model. 
In this approach, the amplitude is written in the Regge 
parametrization as:
\begin{eqnarray}
	{\cal M} &=& \Sigma_{i,j} C_i^2 C_j^2 \left( 
	C_{\Pom} \left( \frac{s}{s_0} \right)^{\alpha_{\Pom}(t)-1} F(t)
	+ C_{\Reg} \left( \frac{s}{s_0} \right)^{\alpha_{\Reg}(t)-1}F(t) \right)
	\; ,
	\nonumber \\
	&+& \Sigma_{i,j} C_i^2 C_j^2 \left( 
	C_{\Pom} \left( \frac{s}{s_0} \right)^{\alpha_{\Pom}(u)-1} F(u)
	+ C_{\Reg} \left( \frac{s}{s_0} \right)^{\alpha_{\Reg}(u)-1} F(u) \right)
	\; .
	\label{SS_amplitude}
\end{eqnarray}
In the simplest version of the model $i, j = \rho^0, \omega, \phi$ 
(only light vector mesons) are included in the standard version.
The couplings $C_i, C_j$ describe the $\gamma \to V_{i/j}$ transitions
that are calculated based on vector meson dilepton width.
$C_{\Pom}$ and $C_{\Reg}$ are extracted from the Regge factorization 
hypothesis (see e.g. \cite{SNS2002}).

It was shown that the component is concentrated mainly at small photon 
transverse momenta which at not too small subsystem
energies corresponds to $z \approx \pm$ 1.
The Regge trajectories are taken in a linear form:
\begin{eqnarray}
\alpha_{\Pom}(t/u) = \alpha_{\Pom}(0) + \alpha_{\Pom}'t/u \; , \nonumber \\
\alpha_{\Reg}(t/u) = \alpha_{\Reg}(0) + \alpha_{\Reg}'t/u \; .
\label{linear_trajectories}
\end{eqnarray}
These linear forms are valid at not too large $|t|$ or $|u|$.
At large $|t|$ or $|u|$ the energy dependent factors are artificially 
small.
Therefore here, where we explore it more, we propose to smoothly
switch off the $t/u$ dependent terms in (\ref{linear_trajectories})
at $t \sim$ -0.5 GeV$^2$ and $u \sim$ -0.5 GeV$^2$.

\begin{figure}
\centering
\includegraphics[width=5cm]{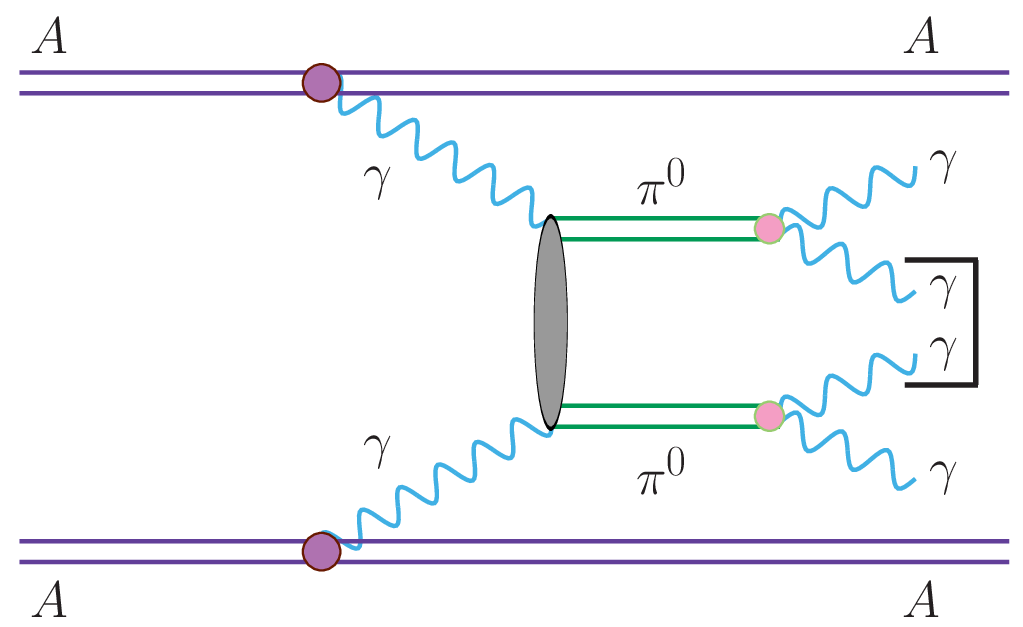}
\caption{Diagram of 2$\pi^0$ photo-production - the dominant background contribution in the region of small di-photon masses.}
\label{fig:background}
\end{figure}

Not only signal but also background contribution shown in
Fig.\ref{fig:background} were included in our analysis.
It was shown that the background contribution becomes important
at smaller $M_{\gamma \gamma} <$  1 GeV \cite{Klusek-Schicker}.

\begin{figure}[!h]
\centering
\includegraphics[width=7cm]{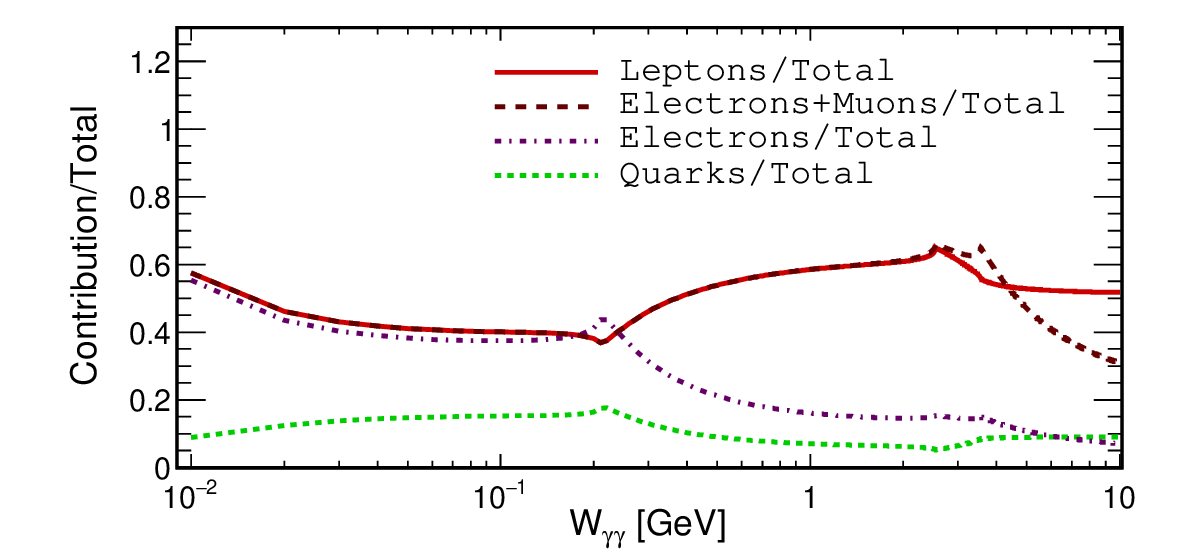}
\caption{A ratio of each contribution to a coherent sum of them. 
	The red solid line represents leptonic cross section divided 
	by the total cross section, the green dotted line refers to 
        quarks, magenta dash-dotted line to electrons and dashed 
        dark-red line applies to a sum of electron and muon 
        contributions.}
\label{fig:interference}
\end{figure}

There are many box diagrams (see diagram (a) in Fig.\ref{fig:diagrams})
with leptons, quarks or $W$ bosons.
Fig.\ref{fig:interference} illustrates how important are interference
effects between the different contributions to the amplitude.

\begin{figure}[!h]
	\centering
(a)\includegraphics[scale=0.23]{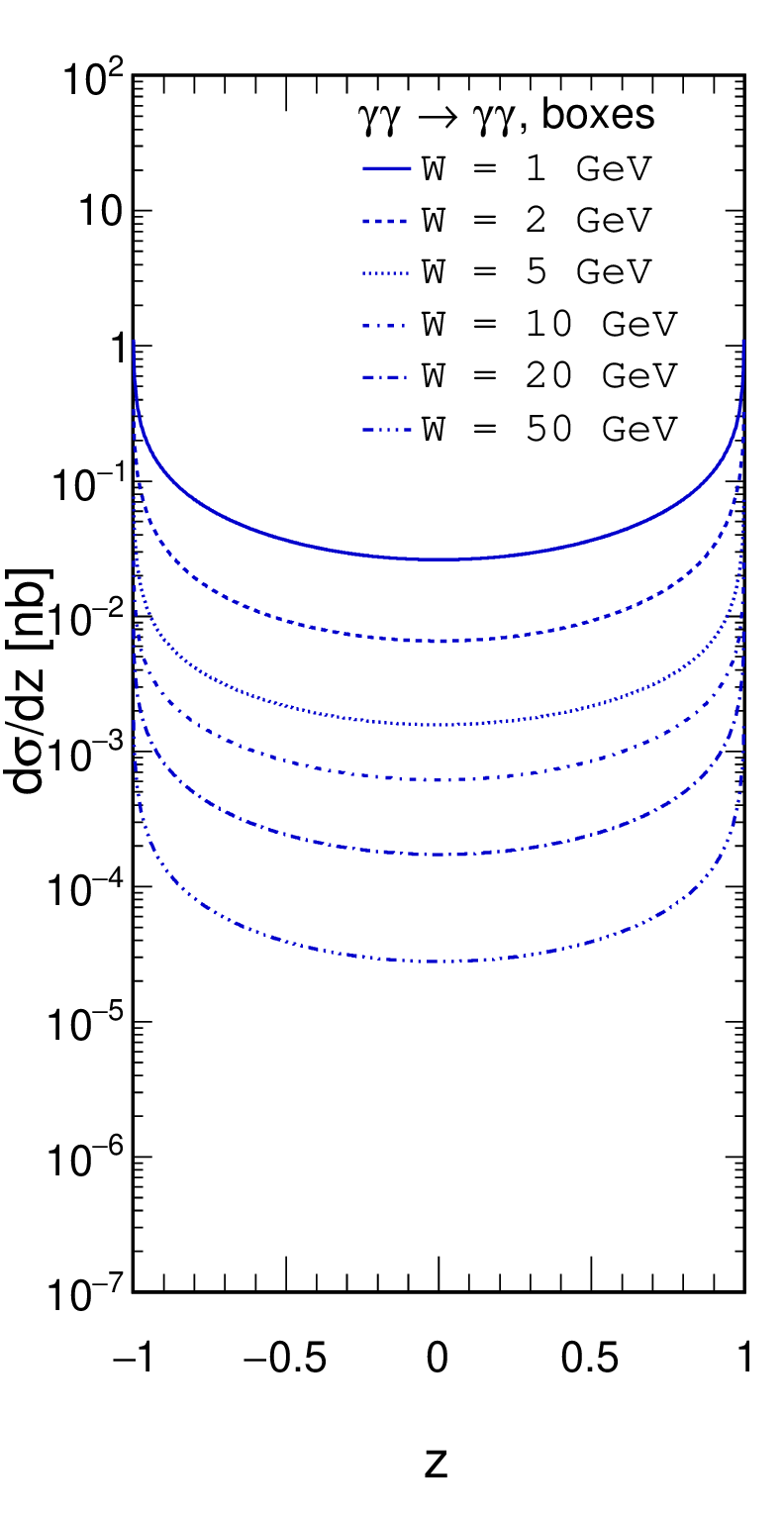}
(b)\includegraphics[scale=0.23]{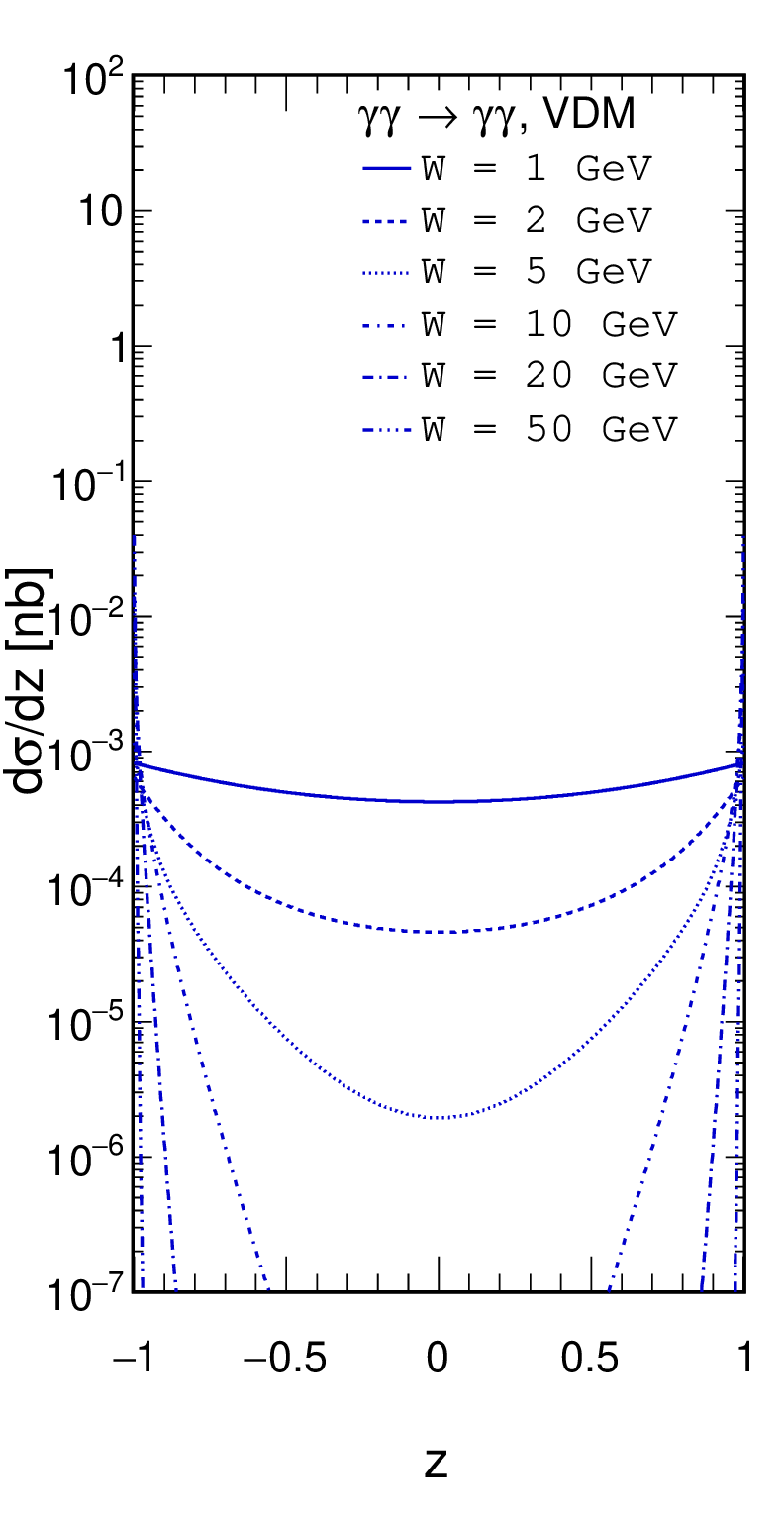}
(c)\includegraphics[scale=0.23]{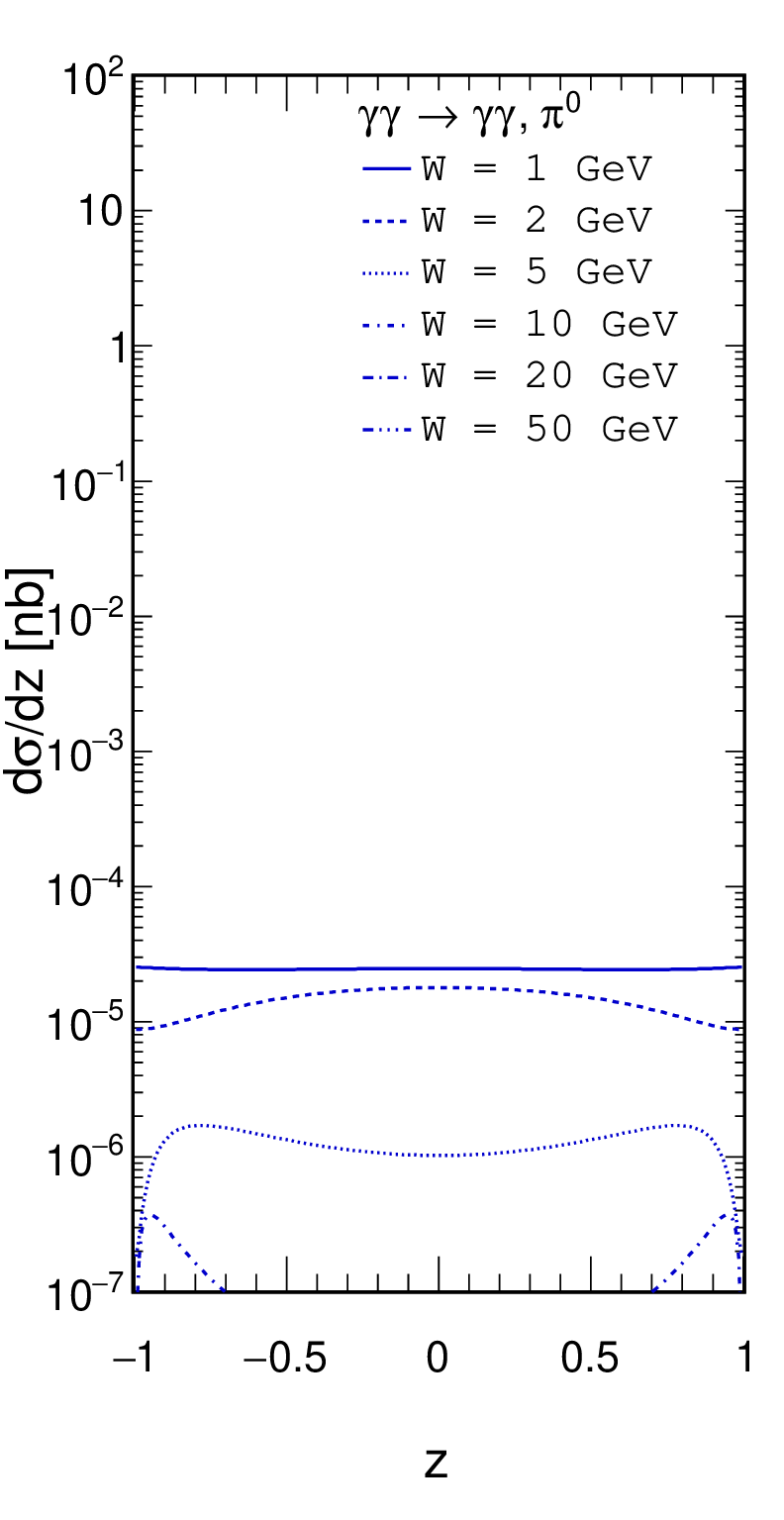}
\caption{$cos(\theta)$ distributions for (a) boxes, 
	(b) double hadronic fluctuations
	and (c) $t/u$-channel $\pi^0$ exchange for different
	energies $W$ = 1, 2, 5, 10, 20, 50$~$GeV.
}
\label{fig:gamgam_gamgam_subleading}
\end{figure}

In Fig.\ref{fig:gamgam_gamgam_subleading} we show angular distributions
for box and some subleading contributions.
The box contribution is by far the largest contribution. The other
contributions are much smaller. In the talk we also discussed interference
effects between the box and double vector meson fluctuations.
We have found negative interference leading to 10 \% modification 
(reduction) of the box result.

\begin{figure}
\centering
\includegraphics[scale=0.28]{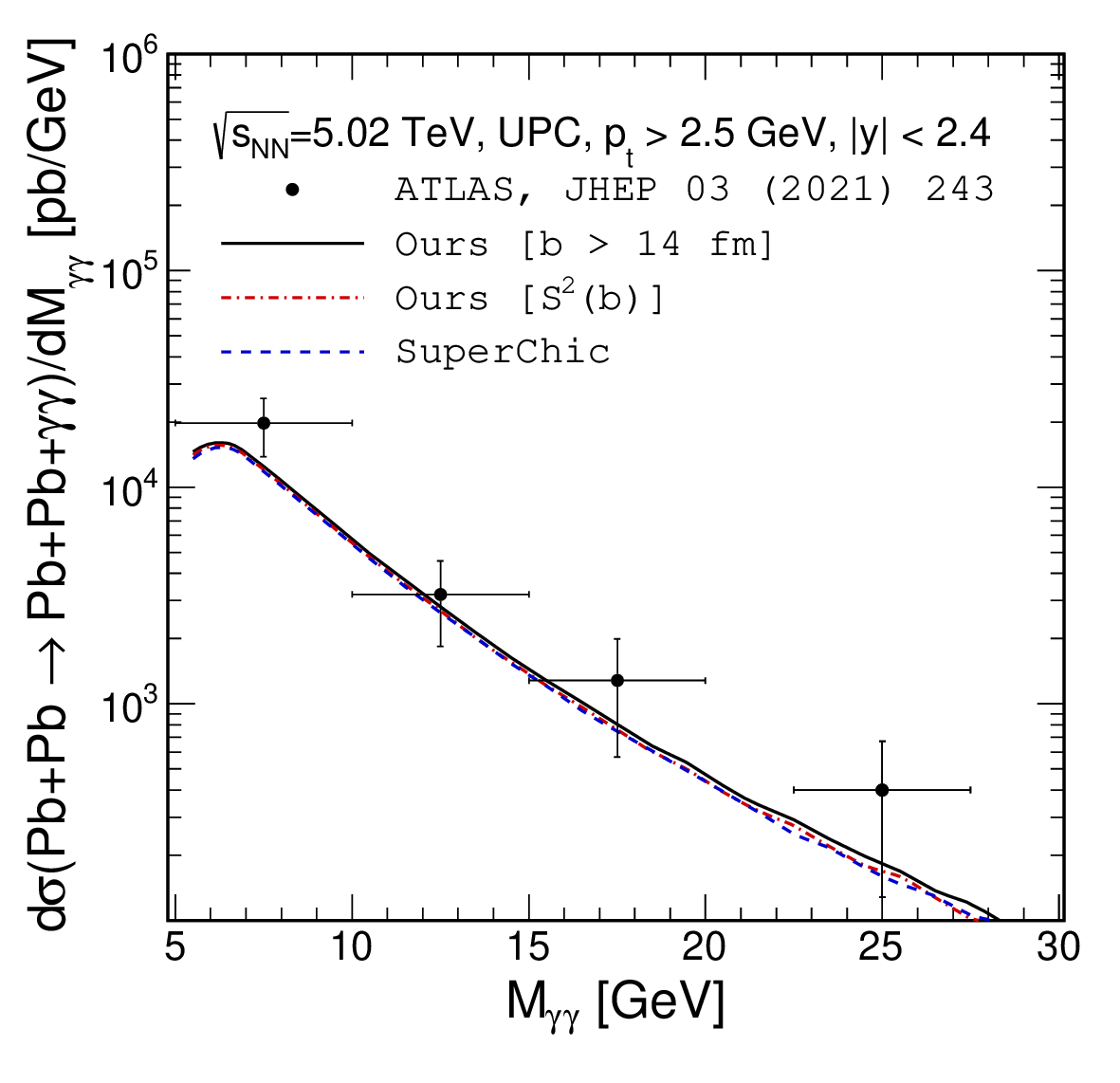}
\caption{\label{fig:2} Differential cross section as a function 
	of two-photon invariant mass at $\sqrt{s_{NN}}=5.02$~TeV.
	The ATLAS data \cite{ATLAS} are shown for comparison.}
\label{fig:ATLAS}
\end{figure}

The box mechanism approximately describes the ATLAS and CMS data
but a more precise comparison to the ATLAS data (see e.g.
Fig.\ref{fig:ATLAS}) shows some missing strength.

What is the missing strength is not fully understood.
Some authors \cite{Biloshytskyi} suggested that the $X(6900)$ fully charmed 
tetraquark could be responsible for the missing strength. However, 
a more refined calculation presented at this conference suggests that it is not
the right explanation \cite{Longjie}.

\section{Inelastic contributions}

Recently in Ref.~\cite{KGS2025} we considered also inelastic
processes when one or both photons couple to nucleons instead
to the whole nucleus. The inelastic contributions (diagrams (b) and 
(c)) were not calculated before.

\begin{figure}[!h]
\centering
\includegraphics[scale=0.20]{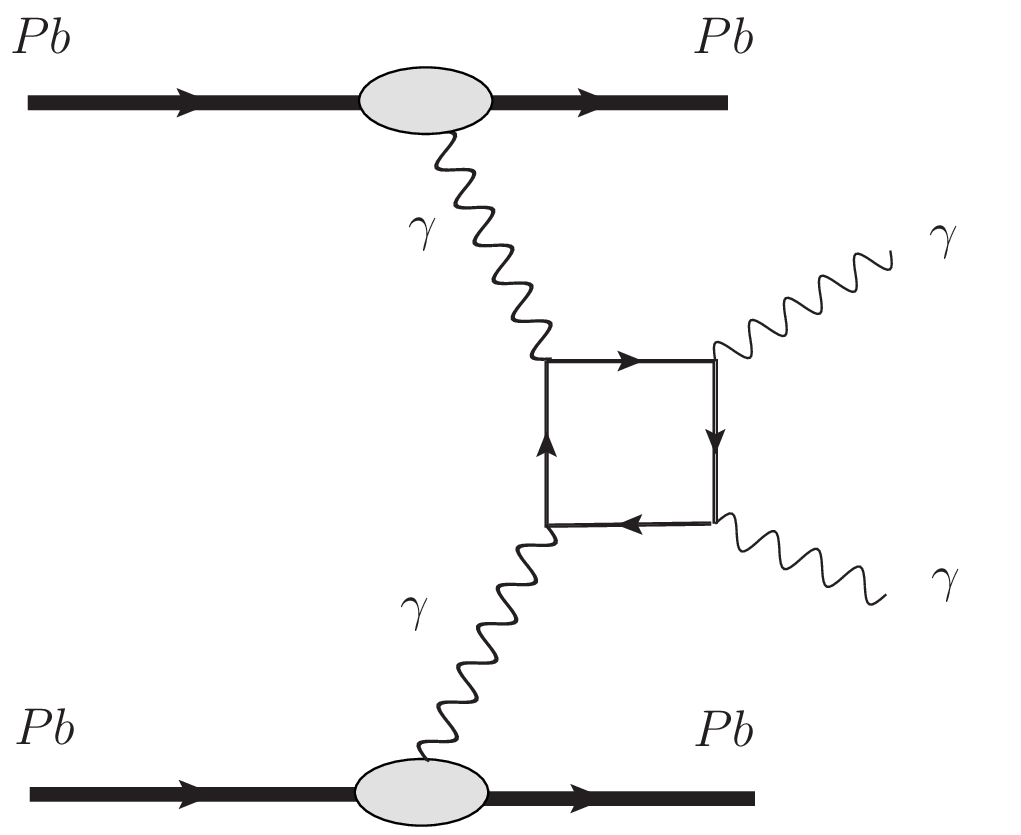}
\includegraphics[scale=0.20]{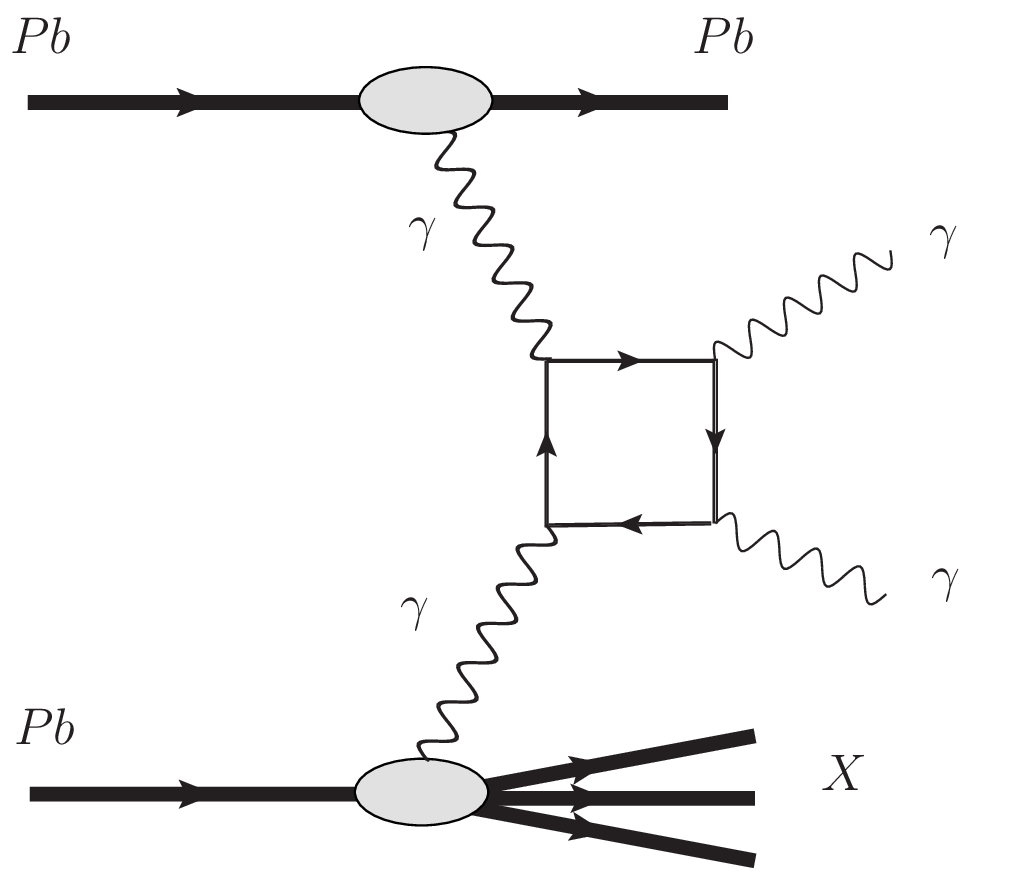}
\includegraphics[scale=0.20]{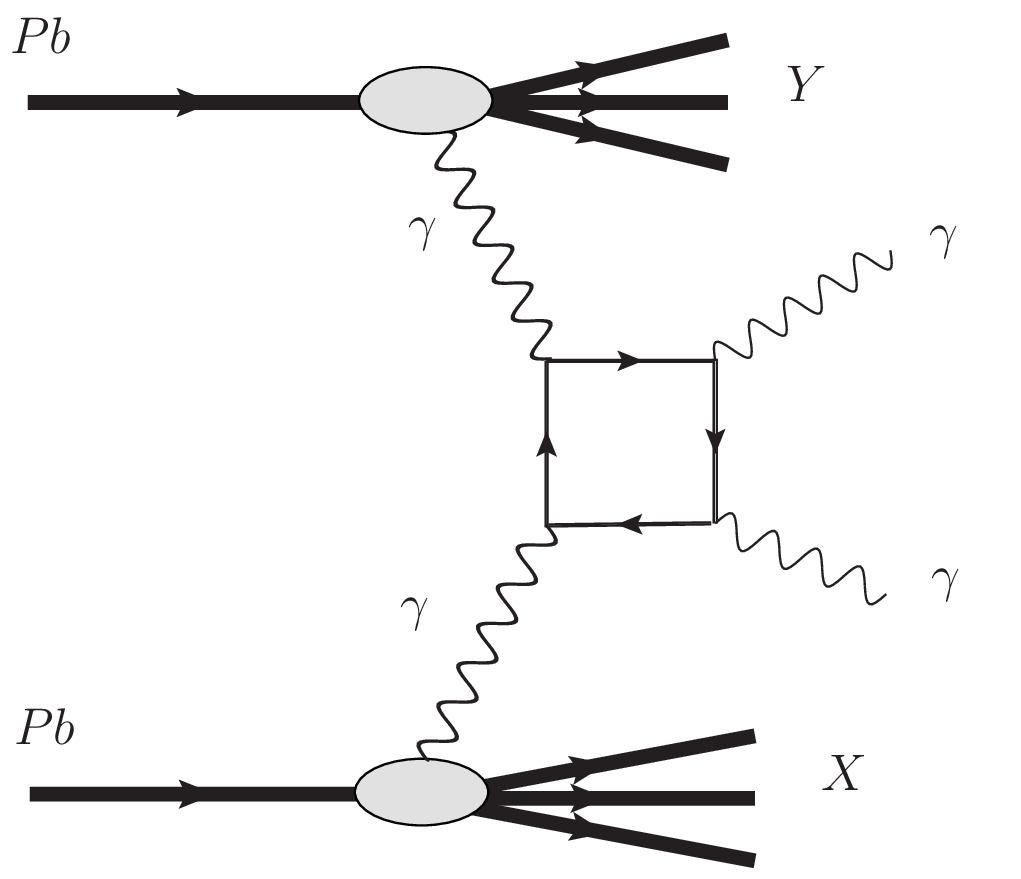}   
\caption{New inelastic mechanisms of diphoton production
	in UPC of heavy ions.}
\end{figure}

The inelastic contributions were not calculated so far.
The inclusion of the inelastic component in the calculation of LbL 
scattering in $PbPb$ collisions, implies that the cross-section 
is given schematically by 
\begin{eqnarray}
\sigma^{LbL} (\sqrt{s_{NN}})   
&\propto&    f_{\gamma/Pb}^{el}(x_1) \otimes f_{\gamma/Pb}^{el}(x_2)  \otimes  \, \hat{\sigma}\left[\gamma \gamma \rightarrow \gamma \gamma ; W_{\gamma \gamma} \right]  \,\,  \nonumber \\
& + & f_{\gamma/Pb}^{el}(x_1) \otimes f_{\gamma/Pb}^{inel}(x_2)  \otimes  \, \hat{\sigma}\left[\gamma \gamma \rightarrow \gamma \gamma ; W_{\gamma \gamma} \right]  \,\,  \nonumber \\
& + & f_{\gamma/Pb}^{inel}(x_1) \otimes f_{\gamma/Pb}^{el}(x_2)  \otimes  \, \hat{\sigma}\left[\gamma \gamma \rightarrow \gamma \gamma ; W_{\gamma \gamma} \right]  \,\,  \nonumber \\
& + & f_{\gamma/Pb}^{inel}(x_1) \otimes f_{\gamma/Pb}^{inel}(x_2)  \otimes  \, \hat{\sigma}\left[\gamma \gamma \rightarrow \gamma \gamma ; W_{\gamma \gamma} \right]   \,\,\, ,
\label{Eq:LbL}
\end{eqnarray}
In a simplified calculation of elastic contribution used for reference (diagram (a)) 
the flux of photons is calculated as
\begin{eqnarray}
f_{\gamma/Pb}^{el} (x) & = & \frac{\alpha Z^2}{\pi x}\left\{2\xi K_0(\xi)K_1(\xi) - \xi^2[K_1^2(\xi) - K_0^2(\xi)] \right\}  \,\,.  
\end{eqnarray}
Photon distributions in proton or neutron were obtained by solving
modified DGLAP equation and were taken from CTEQ18QED 
\cite{Xie:2021equ,Xie:2023qbn}
\begin{eqnarray}
f_{\gamma/Pb}^{inel} (x,\mu^2) & = & 
Z \times f_{\gamma/p} (x,\mu^2) + 
(A - Z) \times f_{\gamma/n} (x,\mu^2) \,\,,
\end{eqnarray}

\begin{figure}[!h]
	\centering
\includegraphics[scale=0.28]{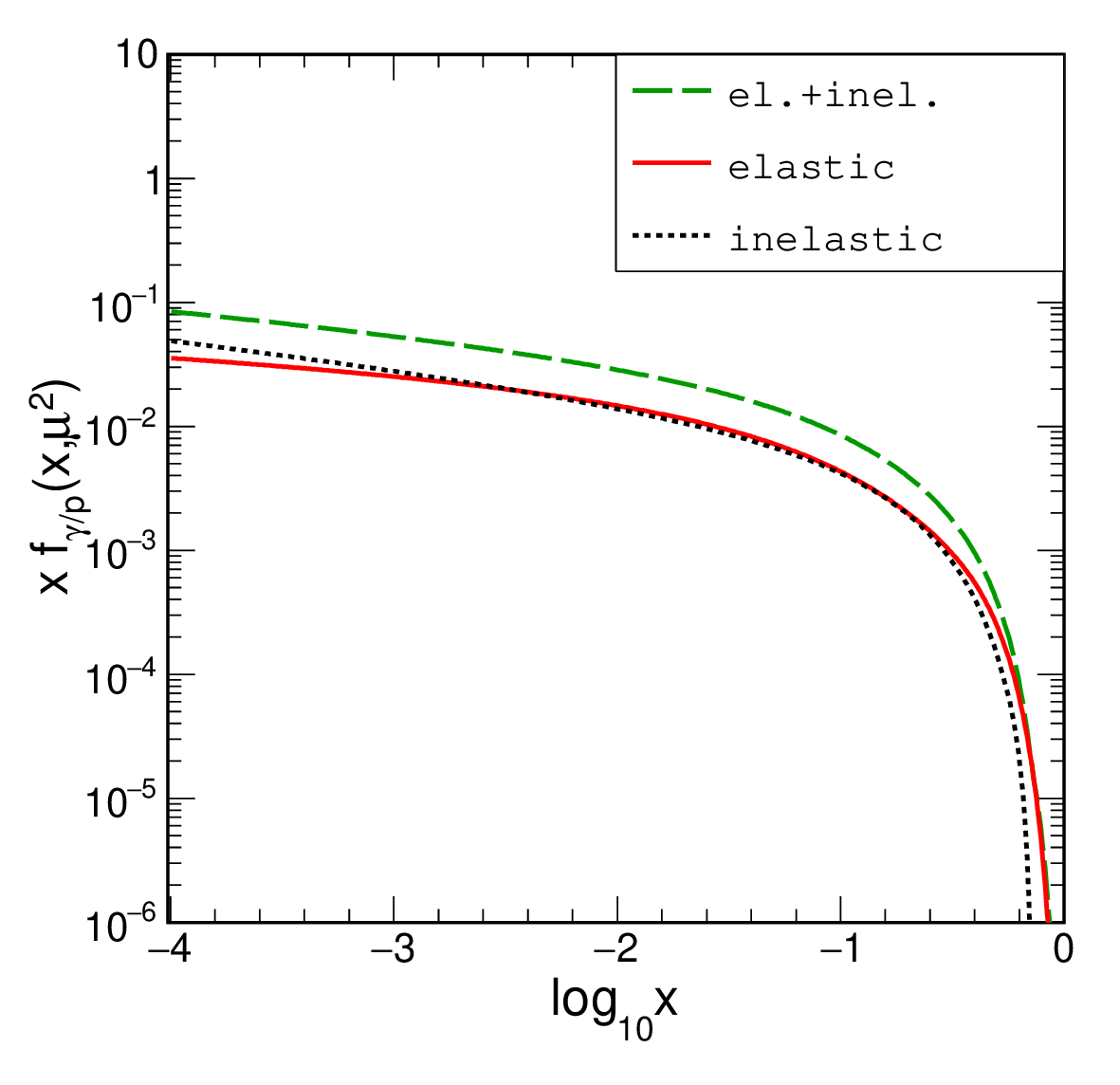}
\includegraphics[scale=0.28]{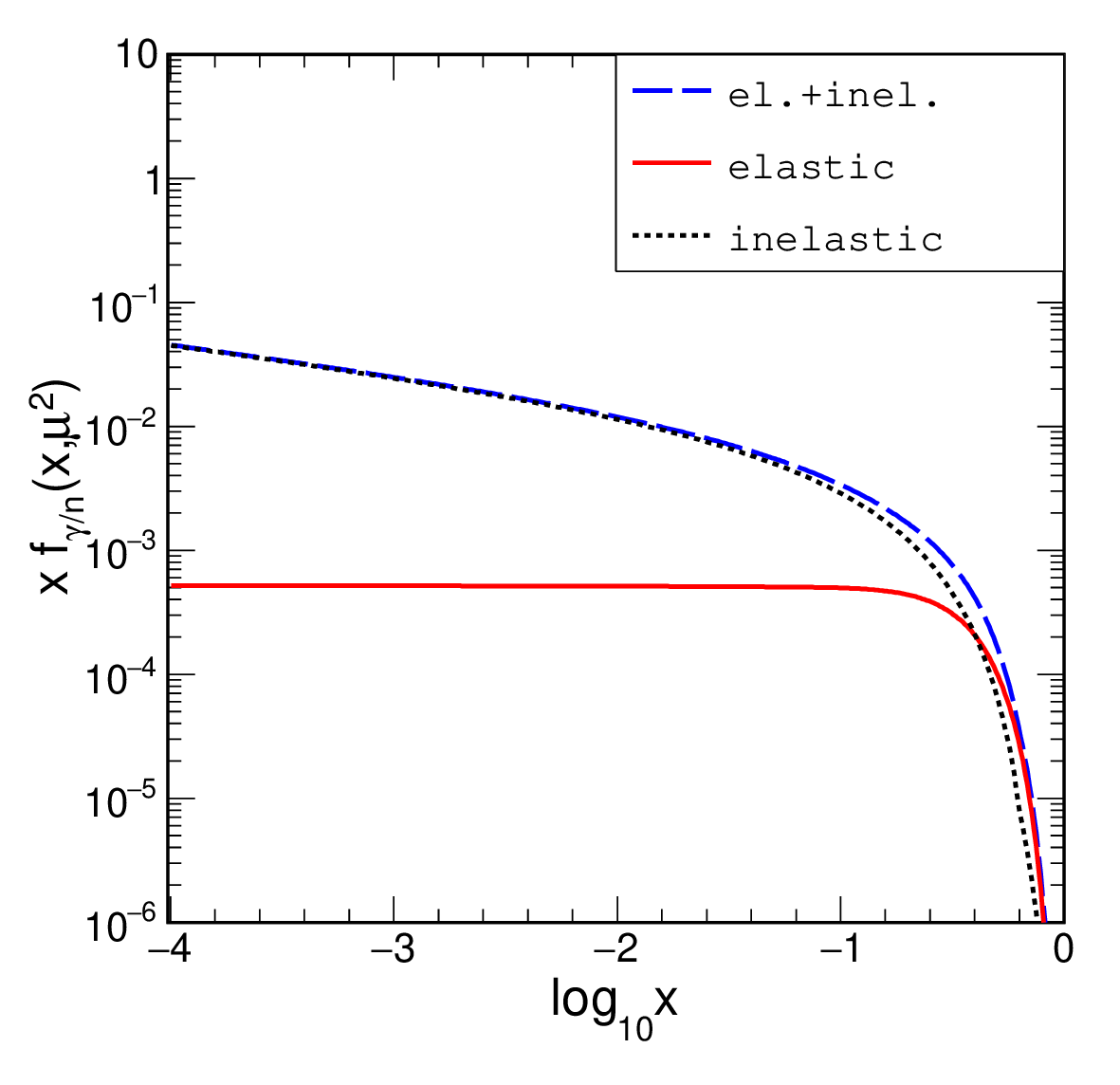}
\caption{$x f_{\gamma/p}(x,\mu^2)$ (left) and
		$x f_{\gamma/n}(x,\mu^2)$ (right).
		Here $\mu$ = 5 GeV.}
	\label{fig:photon_in_nucleons}
\end{figure}

The photon distributions as a function of $log_{10}x$
for fixed factorization scale are shown in 
Fig.\ref{fig:photon_in_nucleons}.
The elastic coupling to nucleons is described in terms of electric
and magnetic form factors. For neutron the electric form factor is zero.
For protons the elastic and inelastic contributions are of similar size,
while for neutrons the elastic part is rather small.

\begin{figure}[!h]
\centering
\includegraphics[scale=0.26]{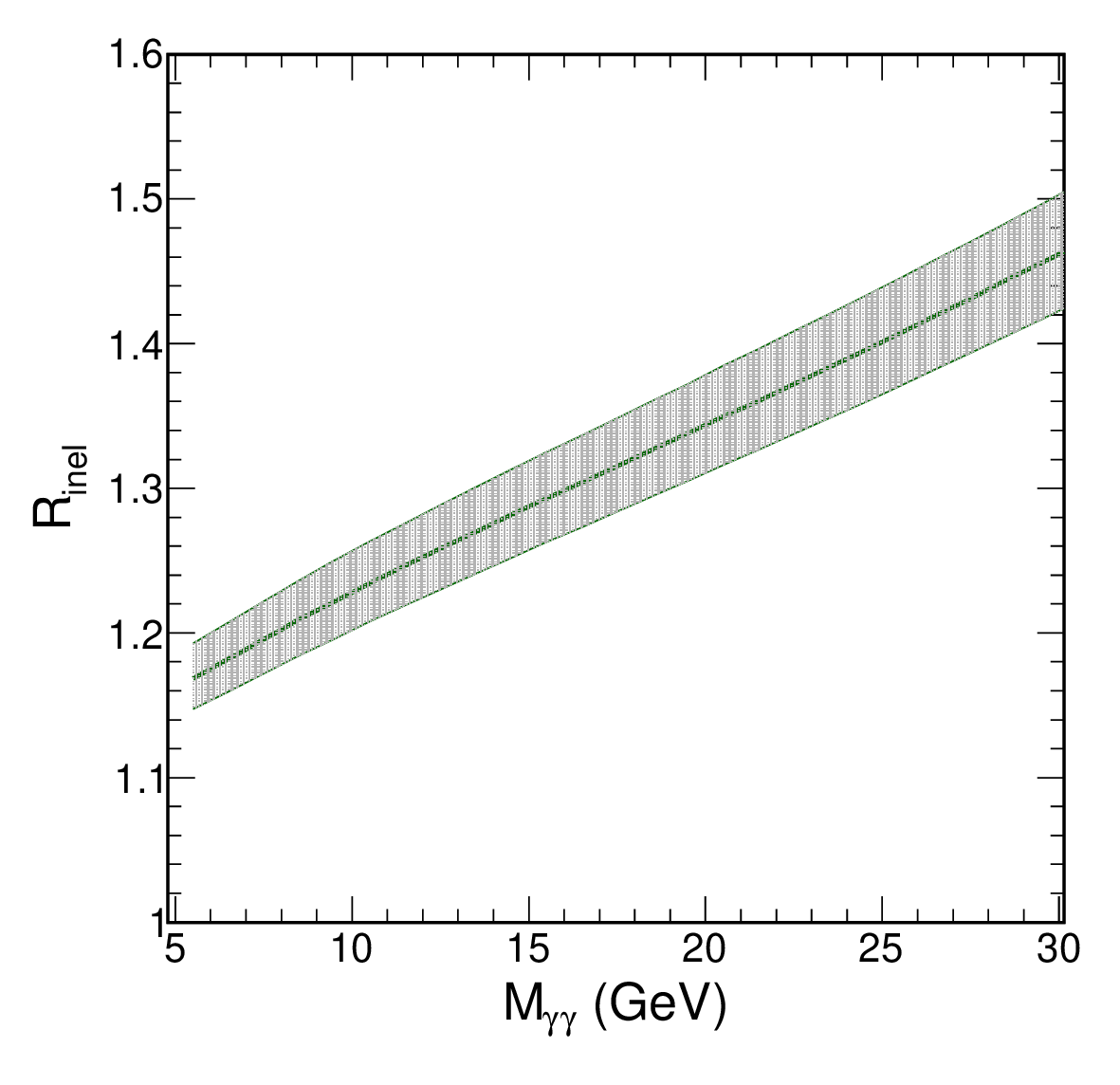}
\includegraphics[scale=0.26]{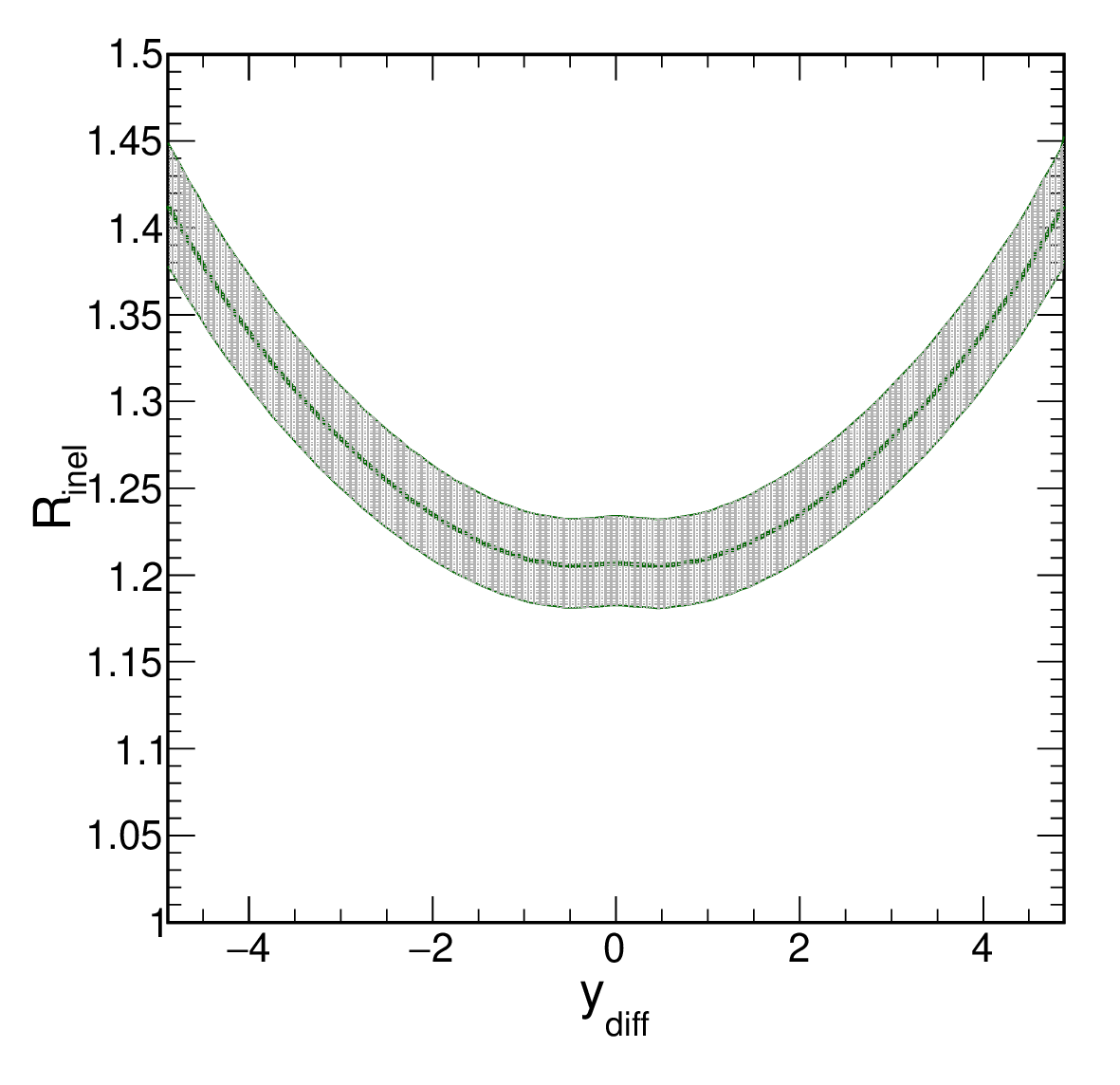}
\caption{The ratio of the modified cross section 
	to the coherent-coherent contribution
	$R_{inel} = \frac{d \sigma^{lbl}(elastic+semi-elastic+inelastic)}
		{d \sigma^{lbl}(elastic)}$,
	$\mu^2 \in (M^2_{\gamma \gamma}/2, 2 M^2_{\gamma \gamma})$.}
\label{fig:inelastic_to_elastic}
\end{figure}

Our inelastic flux does not include shadowing effects.
In addition, it is not clear how the experimental conditions
eliminate the inelastic contributions.
In spite of this in Fig.\ref{fig:inelastic_to_elastic}
we show the ratio of the maximal inelastic-to-elastic contributions
as a function of diphoton invariant mass (left panel) and
rapidity difference (right panel).
As can be seen in the figure, the ratio grows with
invariant mass and rapidity difference.

\begin{figure}[!h]
\centering
\includegraphics[width=6cm]{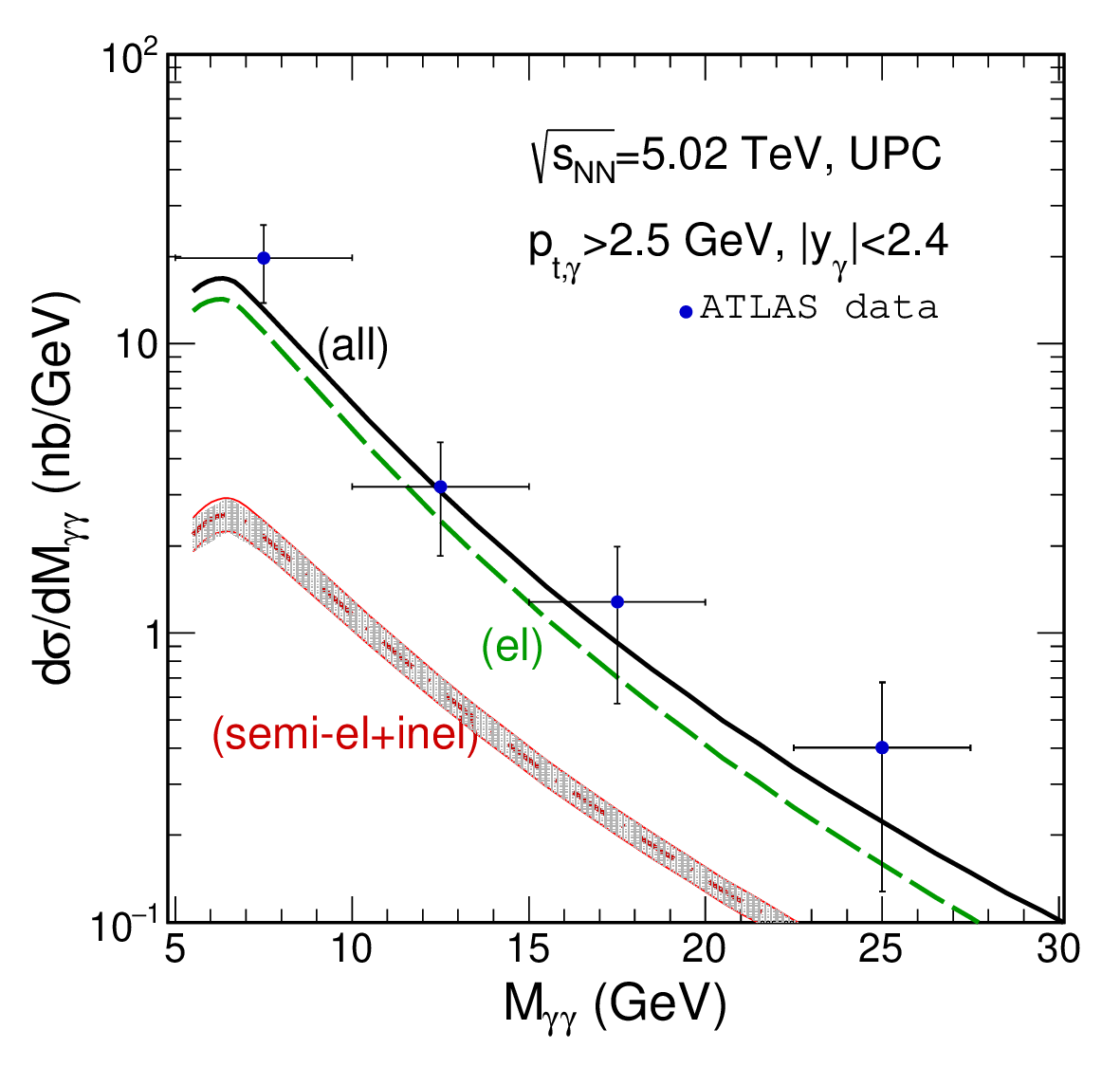}
\caption{Diphoton invariant mass distribution.
	Here for illustration we have added elastic and maximal 
        inelastic contributions which is naive as discussed 
        in the text.}
\label{fig:inelastic_in_ATLAS}
\end{figure}

Fig.\ref{fig:inelastic_in_ATLAS} we show maximal (no extra cuts on 
exclusivity) inelastic contribution together with the ATLAS data. 
For illustration we show also a sum of elastic and maximal inelastic 
contributions. 

Uncertainties illustrated in Fig.\ref{fig:inelastic_to_elastic}
and \ref{fig:inelastic_in_ATLAS} for inelastic contributions are due to
the choice of the scale in parton (photon) distributions that are
varied in the interval 
$\mu^2 \in (M_{\gamma \gamma}^2/2, 2 M_{\gamma \gamma}^2)$.

\section{$\gamma \gamma \to \gamma \gamma$ scattering and
neutron emission}

\begin{figure}[H]
\centering
\includegraphics[width=7cm]{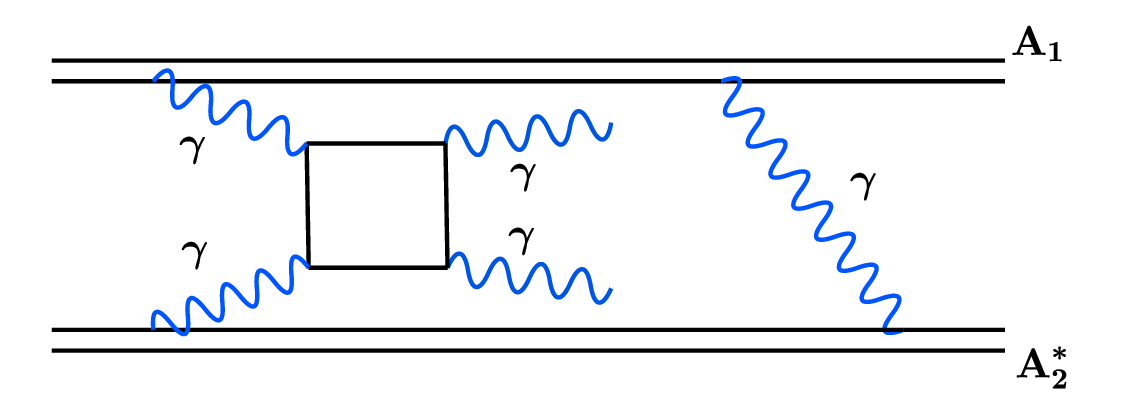}
\caption{A schematic diagram for the $\gamma \gamma \to \gamma \gamma$
scattering in UPC associated with emission of neutron in
the two-step processes.}
\label{fig:associated}
\end{figure}

The photons in UPC can interact with each other, but also lead to excitation
of nuclei participating in the collision. Both processes may occur simultaneously, which is the topic of article \cite{JKS2025}.
In Fig.~\ref{fig:associated} we show a schematic picture
of this coincidence event.

The mathematical approach for description of these events starts from the
mean number of absorbed photons definition:

\begin{equation}
	m(b) = \int d\omega N(\omega, b) \sigma_{abs}(\omega) P_n(\omega) ,
\end{equation} 

\noindent where $N(\omega, b)$ is a photon flux, and $\sigma_{abs}(\omega)$ absorption cross section. The additional 
factor:

\begin{equation}
	P_n(\omega) = 1 - \text{exp}\bigg[\frac{-(\omega-E_{sep})}{\lambda}\bigg],
	\label{factor_prob}
\end{equation}

\noindent was applied, to include the observation made in \cite{GSI} where low-energetic photon absorption does
not always cause the neutron emission. The separation energy in our approach is $E_{sep}$~=~7.37~MeV and $\lambda$~=~0.8~MeV.
The probability of neutron emission from single nucleus and lack of break-up can be denoted respectively as:

\begin{equation}
	\begin{split}
		&P^{Xn}(b) = 1-\text{exp}[-m(b)], \\
		&P^{0n}(b) = \text{exp}[-m(b)].
	\end{split}
\end{equation}

\noindent Both nuclei can be excited in collision, thus the probabilities for different categories can be written

\begin{equation}
	\begin{split}
		&P^{XnXn}(b) = \big(P^{Xn}(b) \big)^2, \\
		&P^{0n0n}(b) = \big(P^{0n}(b) \big)^2, \\
		&P^{Xn0n}(b) = P^{Xn}(b)P^{0n}(b). \\
	\end{split}
\end{equation}

\begin{figure}
\centering
\includegraphics[width=0.46\textwidth]{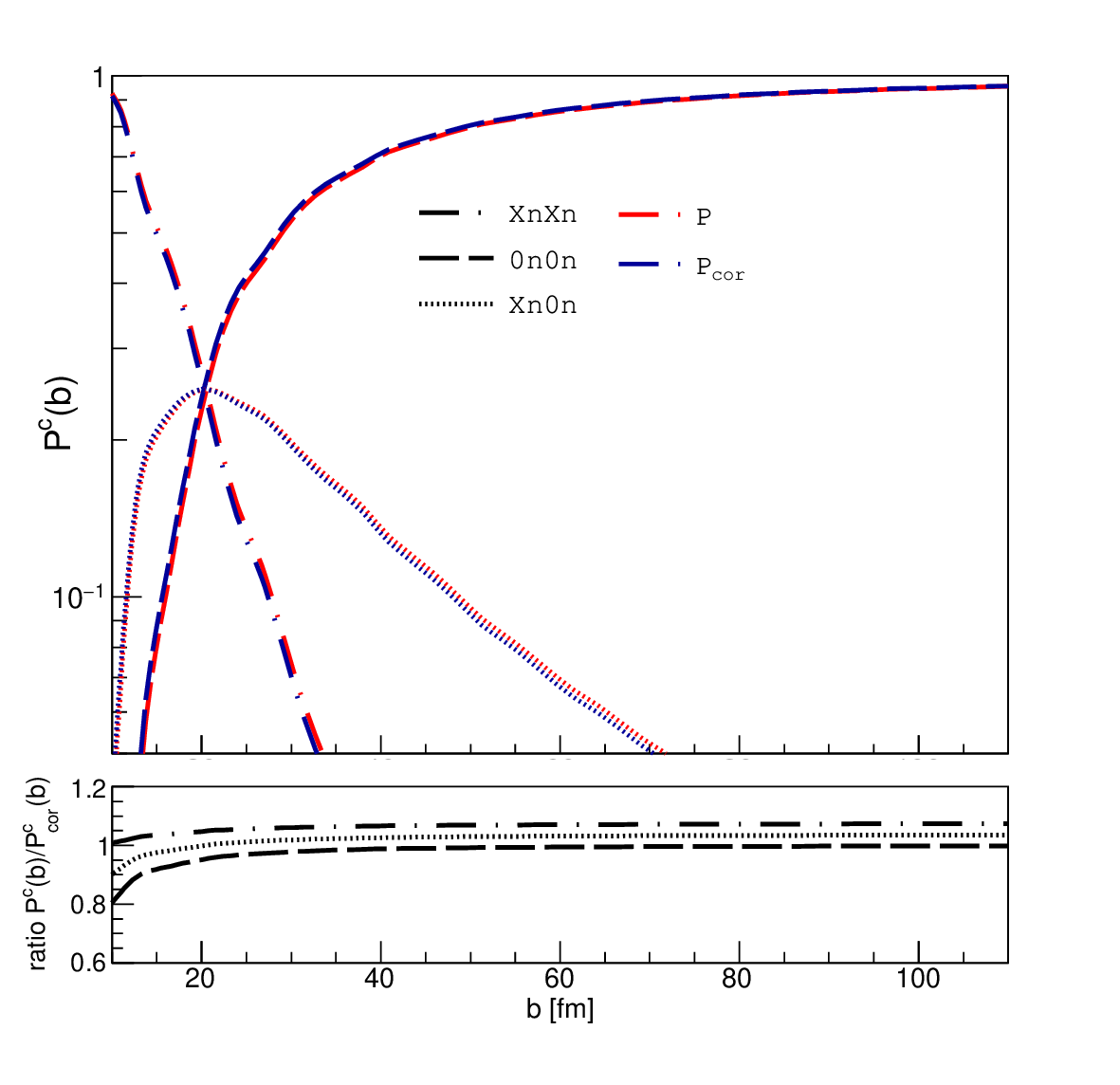}
\caption{Break-up probabilities as a function of impact parameter:
	$XnXn$ (dash-dotted line), $0n0n$ (dashed line) and $Xn0n$ (dotted
	line). The lower panel is explained in the main text. }
\label{fig:probabilities}
\end{figure}

\noindent In Fig.~\ref{fig:probabilities} we show the break-up probabilities for
different neutron categories. There is strong dependence on impact
parameter for different neutron categories. The $P$ and $P_{corr}$ differentiate between
considering the factor given by~(\ref{factor_prob}). Finally, the probability of each category is
implemented to the integral~(\ref{eq:tot_xsec}).

In Table~\ref{tab1} we show a summary of cross sections for 
$Pb Pb \to Pb Pb \gamma \gamma$ for different neutron categories.
Potential deviation from our predictions would be a valuable new 
information.

\begin{table}
\centering
\caption{Cross section in nb for light by light scattering in $Pb+Pb$
	UPC and different neutron categories. Here $\sqrt{s_{NN}} = 5.36$ TeV 
	corresponding to the ongoing ATLAS analysis.}
\label{tab1}
\begin{tabular}{c|c c c c }
	\hline
	&  $\sigma_{total}$         & $0n0n$ & $XnXn$ & $Xn0n+0nXn$ \\ \hline
	cross section [nb]  & 81.886     & 61.679 & 4.261 & 15.963  \\
	percent of $\sigma_{total}$ [\%]     &        &  75.30  & 5.20 &  9.50  \\ \hline
\end{tabular}
\end{table}

\begin{figure}[!h]
\centering
a) \includegraphics[width=0.45\textwidth]{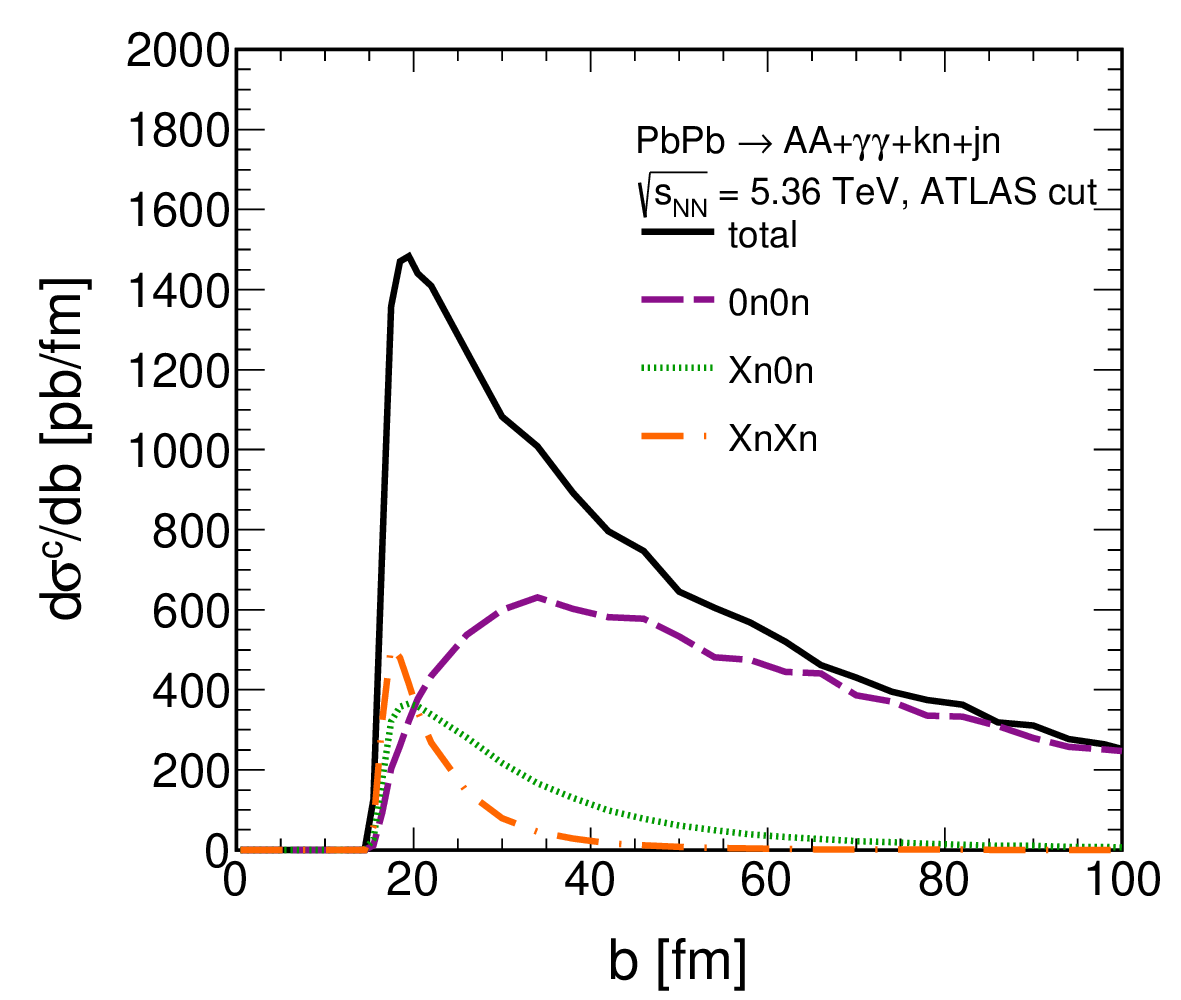}
b) \includegraphics[width=0.45\textwidth]{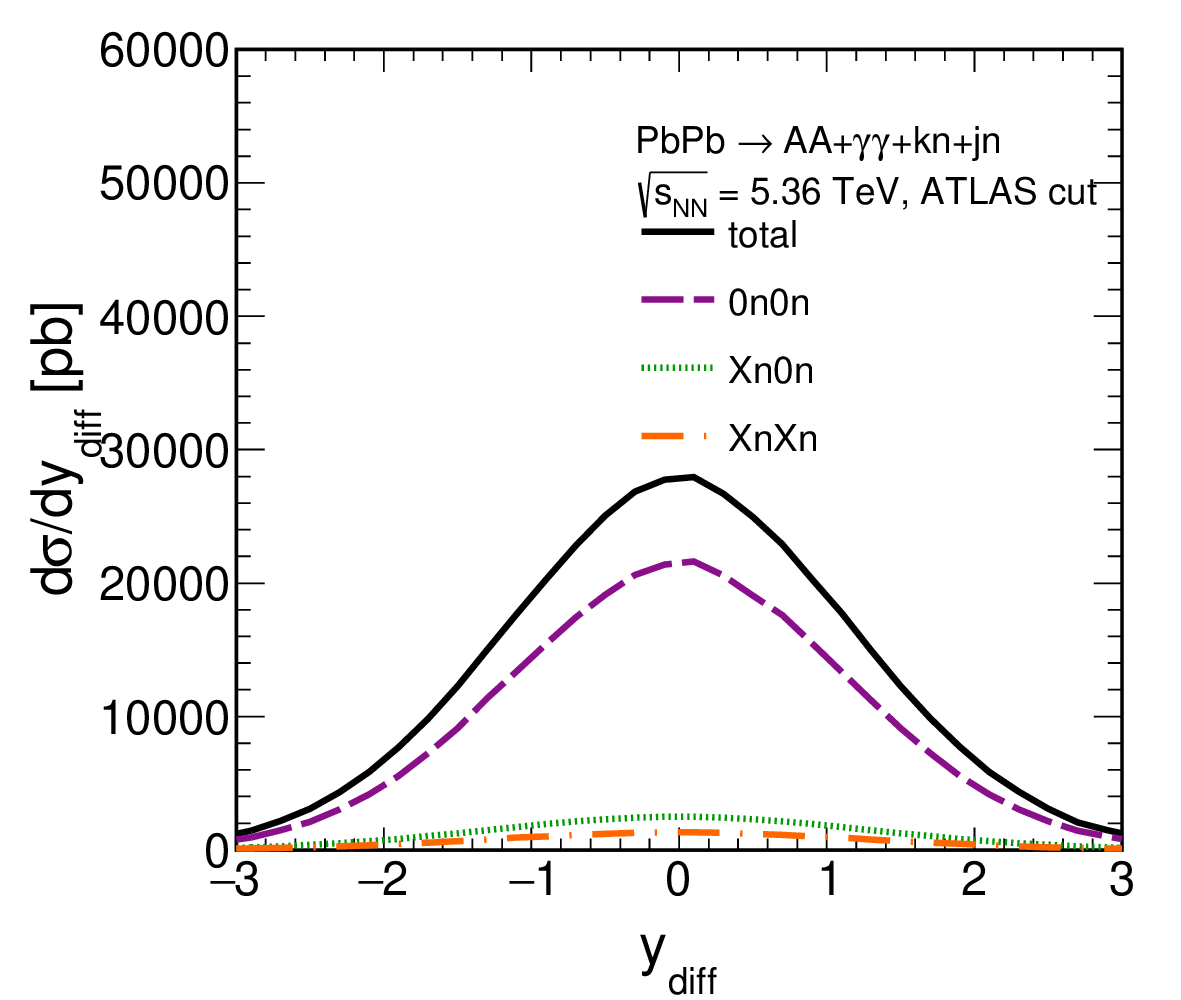}
\caption{a) Impact parameter distribution, and b) Rapidity difference distribution of 
	cross section for: total cross section (solid line), $0n0n$ (dashed line), $Xn0n$ (dotted
	line) and $XnXn$ (dash-dotted line).}
\label{fig:dsig_db}
\end{figure}

In Fig.~\ref{fig:dsig_db}a) we show impact parameter dependence for
the total cross section. With greater distance between nuclei, the nuclear break-up
is less possible. In Fig.\ref{fig:dsig_db}b) we show distribution in $y_{diff} = y_1 - y_2$.
There is only a small dependence of the shape on neutron category \cite{JKS2025}.
This could be confirmed experimentally. A deviation could signal the presence of the inelastic
contributions discussed in the previous section.

\section{Single photon production}

So far we calculated cross section for $A A \to A A \gamma \gamma$
processes.
Finally we wish to show also cross section for single
photon production. 

\begin{figure}
\centering
\includegraphics[width=5cm]{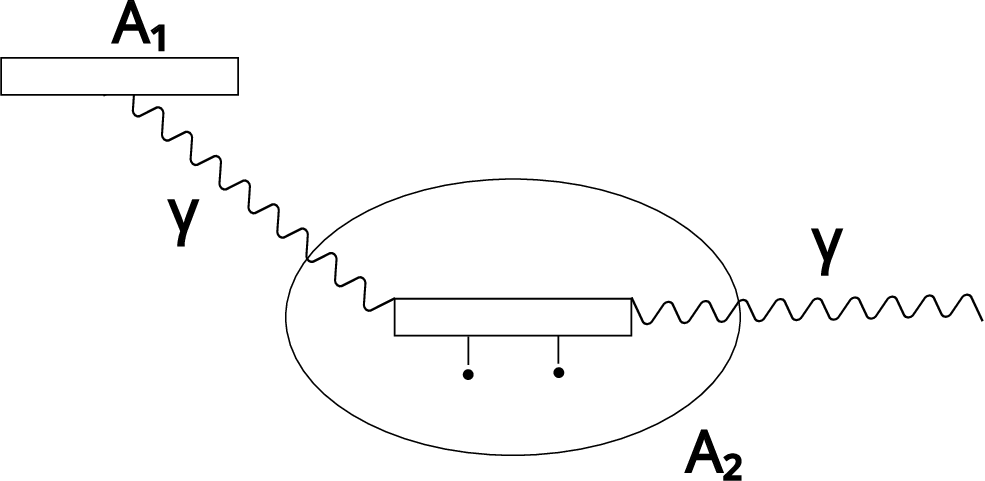}
\includegraphics[width=5cm]{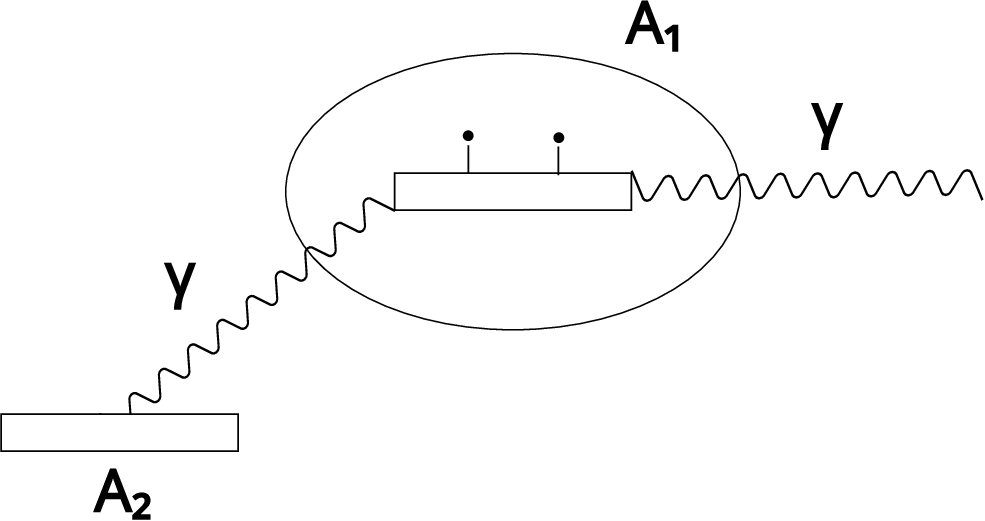}
\caption{Two mechanisms of one photon production based on a
	specific mechanism of Compton scattering in UPC of heavy 
	ions.}
\label{fig:elastic_rescattering_Compton}
\end{figure} 
In the spirit of vector dominance model the differential distribution 
for elastic Compton scattering is related to the similar distribution 
for vector meson scattering as
\begin{eqnarray}
\frac{d \sigma(\gamma A \to \gamma A; t=0)}{dt} =
\sum_V \left(\frac{4 \pi \alpha_{em}}{f_v^2} \right)^2
\frac{d \sigma(\rho^0 A \to \rho^0 A; t=0)}{dt} \; .             
\end{eqnarray}
Here the sum runs over three light vector mesons: 
$V = \rho^0, \omega, \phi$.
The differential distribution at $t$ = 0 can be expressed
in terms of the total cross section for $V+A$ scattering as:
\begin{equation}
\frac{d \sigma(\rho^0 A \to \rho^0 A, t=0)}{dt}
=\frac{1}{16 \pi} \sigma_{tot}^2(\rho^0 A) F^2(t) \; .
\end{equation}

Then the total cross section for $\rho^0 A$ scattering can
be calculated in the Glauber model as:
\begin{equation}
\sigma_{tot}(\rho^0 A; k) = \int d^2 \rho
\left( 1 - \exp(-\sigma_{\rho N}^{tot}(k) T_A(\rho)) \right) \; .
\end{equation}

\begin{figure}[H]
	\centering
	\includegraphics[width=5.25cm]{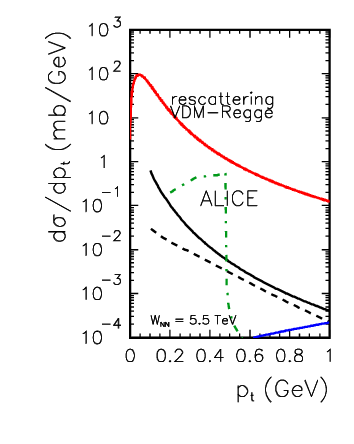}
	\includegraphics[width=5.25cm]{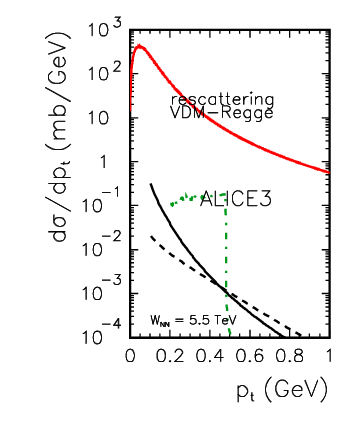}
	\caption{Transverse momentum distribution of photon with one photon
		registration for ALICE and ALICE-3 acceptance for $\sqrt{s_{NN}}$ = 5.5 TeV. 
		The black solid line is for the two-photon box component, 
		the black dashed line is
		for double photon fluctuation into light vector mesons 
		while the red solid line for the genuine single photon production 
		(rescattering effect only).}
	\label{fig:dsig_dpt_ALICE_vs_ALICE3}
\end{figure}

In Fig.\ref{fig:dsig_dpt_ALICE_vs_ALICE3} we show preliminary 
transverse momentum  distributions of single photon for ALICE (left) and ALICE 3 (right)
experiments. We show also distributions for $A A \to A A \gamma \gamma$
when one photon is measured and second is not.

\section{Summary}

In this paper, we have discussed several electromagnetic processes, with a focus on light-by-light scattering. We have presented the main theoretical approach, mechanisms involved, and the current state of experimental knowledge. We have also presented the main arguments in the debate about the missing strength in the ATLAS measurement as a result of neglecting the impact of charm tetraquark contribution.

In the next section, we have summarized our approach to inelastic scattering of $\gamma\gamma\to\gamma\gamma$. In this mechanism, at least one photon couples to the nucleon. Results predict the strength of the process between 20-40\% of the total elastic contribution, and show dependence on kinematics. The uncertainties of our approach come from the choice of factorization scale. We have presented also still open issues:

\begin{itemize}
	\item Modelling of the final state (associated emission of neutrons and protons) excited by truly inelastic processes.
	\item Can one distinguish the final state excited by extra photon exchange from that due to one-step inelastic processes?
	\item Can the inelastic $\gamma \gamma \to \gamma \gamma$ processes be measured? This requires extra studies, including modeling of the gap survival factor.
	\item In the collinear approach to inelastic processes $p_t(\gamma \gamma) =$ 0, but including $k_t$'s of photons $p_t(\gamma \gamma) \ne$ 0. A study with varying cuts on $p_t(\gamma \gamma)$ or accoplanarity would be useful.
\end{itemize}

We have considered in detail the production of neutrons associated with diphoton scattering for future ATLAS analysis. We have calculated fractions of cross section for $\gamma \gamma$ central production associated with different neutron categories. The open question is whether the measurement of neutrons in ZDC could help in identifying the inelastic processes? This is the case for $J/\psi$ production \cite{GSZ2014}.

Finally, we presented our preliminary results on single photon production. Two mechanisms have been discussed:\\
(a) rescattering, as for vector meson production \\
(b) Delbr\"uck scattering (not discussed here)\\
We have also considered two-photon final states when only one of the photons is measured. In this context, we have considered box mechanisms and double photon fluctuations into light vector mesons. The single photon rescattering mechanism dominates over the incompletely measured two-photon final state. ALICE, LHCb, and ALICE-3 could try to measure the single photon production.


\end{document}